\documentclass[11pt,showpacs,preprintnumbers,amsmath,amssymb,prd,nofootinbib,superscriptaddress]{revtex4-2}

\usepackage{dcolumn}
\usepackage{bm}
\usepackage{ifpdf}
\usepackage{hyperref}
\usepackage{xcolor,color,graphicx,graphics,physics}
\usepackage[spanish,english]{babel}
\usepackage[latin1]{inputenc}
\usepackage[OT1]{fontenc}
\usepackage{latexsym,amssymb,amsmath,amsfonts, slashed,cancel}
\usepackage{makeidx}
\usepackage{epsfig,subfigure}
\usepackage{natbib}
\usepackage{epstopdf}
\usepackage{mathrsfs}
\usepackage{hyperref}
\hypersetup{colorlinks=true, linkcolor=blue, citecolor=blue, urlcolor=blue}
\usepackage{enumerate}

\usepackage{fixmath}

\everymath{\displaystyle}
\usepackage{graphicx}

\usepackage[T1]{fontenc}
\usepackage{amsmath}
\usepackage{amssymb}
\usepackage{graphicx}
\usepackage{xcolor}

\newcommand{\bea}{\begin{eqnarray}}
\newcommand{\eea}{\end{eqnarray}}

\newcommand{\orcid}[1]{\href{https://orcid.org/#1}{\includegraphics[width=10pt]{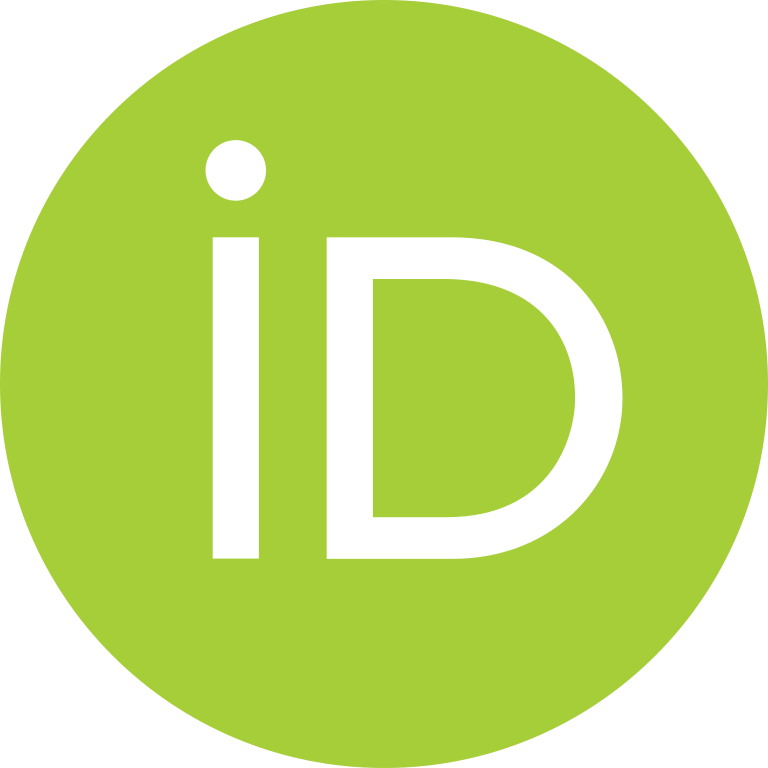}}}

\begin{document}

\title{Thermal Bhabha scattering under the influence of non-hermiticity effects}

\author{D. S. Cabral  \orcid{0000-0002-7086-5582}}
\email{danielcabral@fisica.ufmt.br}
\affiliation{Programa de P\'{o}s-Gradua\c{c}\~{a}o em F\'{\i}sica, Instituto de F\'{\i}sica,\\ 
Universidade Federal de Mato Grosso, Cuiab\'{a}, Brasil}

\author{A. F. Santos \orcid{0000-0002-2505-5273}}
\email{alesandroferreira@fisica.ufmt.br}
\affiliation{Programa de P\'{o}s-Gradua\c{c}\~{a}o em F\'{\i}sica, Instituto de F\'{\i}sica,\\ 
Universidade Federal de Mato Grosso, Cuiab\'{a}, Brasil}

\author{R. Bufalo \orcid{0000-0003-1879-1560}}
\email{rodrigo.bufalo@ufla.br}
\affiliation{Departamento de F\'{\i}sica, Universidade Federal de Lavras,\\
37203-202, Lavras, Minas Gerais, Brazil}

\begin{abstract}
In this paper, we investigate the Bhabha scattering process within the framework of non-Hermitian QED at finite temperature.
In this theory, the hermiticity condition, typically required in quantum field theory to ensure the reality of physical observables, is replaced by the condition of unbroken $PT$-symmetry which favors the introduction of an axial mass and a vector-axial gauge coupling.
Using the Thermofield Dynamics formalism, we derive and comprehensively analyze the thermal differential cross section for the Bhabha scattering.
Furthermore, we explore the high-energy limit of the scattering amplitude and establish constraints upon the axial coupling constant, offering valuable insights into the system's behavior under extreme conditions.
\end{abstract}

\maketitle

\section{Introduction}

Due to the computational simplicity and precision data of the Bhabha scattering (the process $e^+e^- \to e^+e^-$ ), as highlighted in many standard textbooks \cite{peskin, greiner}, makes it an ideal case for theoretical studies in order to set constraints upon physics beyond the standard model \cite{Charneski:2012py,Casana:2012vu,Bufalo:2015eia,deBrito:2016zav}.
Moreover, the process serves as a crucial experimental tool for testing the validity and precision of QED, further emphasizing its importance in high-energy physics.
In this work, the primary objective is to examine thermal properties of the Bhabha scattering within the framework of a non-Hermitian extension of QED.

Quantum field theories (QFTs) are traditionally constructed under the principles of hermiticity, locality, Lorentz symmetry, and CPT invariance \footnote{CPT refers to the discrete symmetries of charge conjugation (C), parity (P), and time reversal (T).}, which are often considered essential conditions.
However, unlike the other principles, hermiticity is not a fundamental physical requirement but rather a mathematical convenience to ensure real eigenvalues for observables.
In fact, there is no intrinsic physical motivation that requires hermiticity, and recent studies suggest that non-Hermitian extensions of QM and QFT can provide a consistent framework for exploring novel phenomena beyond conventional physics \cite{Bender:1998ke,Mostafazadeh:2001jk,Bender:2005tb,Bender:2019,Alexandre:2023afi,Simon:2018zrj,novitsky,Pekduran:2025ljx}. In this framework, the requirement of hermiticity is replaced by the physical condition that an unbroken $PT$-symmetry ensures observable (real) energies.
Actually, this type of models possess the so-called exceptional points, which establish the boundaries between the regimes of broken and unbroken $PT$-symmetry, and which cannot be reproduced by hermitian theories.
This paradigm shift has opened new avenues for exploring non-Hermitian field theories with  $PT$-symmetry, leading to intriguing applications across various domains.

Recently, some interesting points were considered about formal properties of non-Hermitian field theories \cite{Alexandre:2022uns,Sablevice:2023odu,Li:2022prn,Khan:2024mhc,Chernodub:2024lkr}, non-Hermitian Yukawa theories \cite{Alexandre:2015kra,Alexandre:2020bet,Mavromatos:2022heh}, non-Abelian gauge theory \cite{Fring:2019xgw,Fring:2020bvr}, strongly interacting fermion systems \cite{Felski:2022dsx,Mintz:2024soo}, non-Hermitian holographic  setting \cite{Arean:2019pom,Arean:2024lzz,Arean:2024gks}, among others.
Several studies have investigated non-Hermitian QED in this context, highlighting its potential to describe novel physical phenomena and extend the boundaries of traditional quantum field theories \cite{Bender:2006fe,Makris_2008,Alexandre:2015kra, longhi,Ashida:2017rso,El-Ganainy:2018ksn,Midya_2018,Matsumoto:2019are,Ashida:2020dkc, Rodrigo,Cabral:2025mjf}. 
Hence, in order to examine new features of non-Hermitian QED our primary objective is to investigate the thermal Bhabha scattering within this framework.

Bhabha scattering at finite temperature has been explored in various contexts \cite{afsantos1, afsantos2, afsantos3} to study features of physics beyond the Standard Model.
To address field theories at finite temperature, several formalisms have been developed, primarily based on real-time and imaginary-time transformations \cite{khannatfd, temp00, temp1, temp2, temp3}.
For tree-level processes, the Thermofield Dynamics (TFD) real-time formalism is particularly advantageous, since it provides a straightforward framework to incorporate a heat bath in a way that scattering amplitudes can be calculated similarly to $T=0$ methods \cite{khannatfd, temp000, scatter2}.

The TFD formalism is defined by the introduction of a thermal space, constructed as the direct sum of the conventional Hilbert space and an identical copy.
This additional space acts as a heat bath, ensuring that the system remains in thermal equilibrium.
The doubling of degrees of freedom inherent to this framework results in two components for the field propagator: one corresponding to the conventional zero-temperature contribution and the other accounting for thermal corrections.
Alongside the duplication of the Hilbert space, the Bogoliubov transformation plays a central role in this approach, allowing proper inclusion of thermal effects by establishing the respective relation among the thermal and $T=0$ field operators.

This work is organized as follows: Section \ref{prelim} provides preliminary discussions, introducing the framework of non-Hermitian QED (NH QED), free field solutions, and key features, as detailed in Subsection \ref{nonhermitian}. 
Moreover, subsection \ref{tfd} establishes the dynamics of thermal fermionic quantities, including the bosonic propagator at finite temperature.
In Section \ref{scattering}, we focus on evaluate the thermal Bhabha scattering process, presenting the calculations of the differential cross section using modified Feynman rules at the tree level.
To validate our findings and determine physical meaningful quantities, comparisons with experimental data are also included. Finally, the conclusions are summarized in Section \ref{secconclusion}. For completeness, Appendix \ref{app} provides auxiliary functions that ensure a consistent flow of information throughout the paper.

\section{Preliminaries of the Scattering Process}
\label{prelim}

In this section, we present the preliminary discussions necessary to examine the thermal aspects of the Bhabha scattering process within the non-Hermitian QED.
We begin by outlining key details of the non-Hermitian formulation of QED, followed by a brief introduction to the TFD formalism.

\subsection{Explicit solution for non-hermitian QED}
\label{nonhermitian}

Let us begin with the non-Hermitian QED Lagrangian \cite{Alexandre:2015kra}
\begin{eqnarray} \label{lagran}
\mathcal{L}&=&\overline{\psi}\left[i\gamma^\nu\partial_\nu-m-\mu\gamma^{5}\right]\psi-\frac{1}{4}F_{\mu\nu}F^{\mu\nu}+\mathcal{L}_I,
\end{eqnarray}
where the first term represents the free theory of fermions with vector and axial masses, $m$ and $\mu$, respectively. The second term describes photons, with $F_{\mu\nu}=\partial_\mu A_\nu-\partial_\nu A_\mu$ as usual. Finally, the third term corresponds to the interaction contribution, given by
\begin{eqnarray}
\mathcal{L}_I=\overline{\psi}\left[\gamma^\mu A_\mu\left(g_{v}+g_{a}\gamma^{5}\right)\right]\psi,
\end{eqnarray}
which implies the presence of a chiral current with coupling strength $g_a$, in addition to the electric current associated with $g_v$.
A detailed account on the masses in this model in terms of non-hermitian Higgs mechanism and the axial-vector coupling is give in ref. 

\cite{Cabral:2025mjf}.

Moreover, the non-Hermitian Dirac free Lagrangian leads to field solutions that can be expanded in the standard form as
\begin{eqnarray}
\psi\left(x\right) & = & \sum_{s}\int\frac{\text{d}^{3}p}{\left(2\pi\right)^{3}}\sqrt{\frac{M}{E_{p}}}\left[a_{p}^{s}u^{s}\left(p\right)e^{-ip\cdot x}+(b_{p}^{s})^\dagger\upsilon^{s}\left(p\right)e^{ip\cdot x}\right],\\
\overline{\psi}\left(x\right) & = & \sum_{s}\int\frac{\text{d}^{3}p}{\left(2\pi\right)^{3}}\sqrt{\frac{M}{E_{p}}}\left[b_{p}^{s}\overline{\upsilon}^{s}\left(p\right)e^{-ip\cdot x}+(a_{p}^{s})^{\dagger}\overline{u}^{s}\left(p\right)e^{ip\cdot x}\right],
\end{eqnarray}
whose dispersion relation yields the correct electronic mass, specifically the effective mass given by $M^2=m^2-\mu^2$. The fermionic creation and annihilation operators satisfy the following commutation relations
\begin{align}
\left\{ a_{p}^{r},(a_{q}^{s})^{\dagger}\right\}  & =\left\{ b_{p}^{r},(b_{q}^{s})^{\dagger}\right\} =(2\pi)^{3}\delta^{3}(\vec{p}-\vec{q})\delta_{rs},\\
\left\{ a_{p}^{s},(b_{q}^{s})^{\dagger}\right\}  & =\left\{ b_{p}^{s},(a_{q}^{s})^{\dagger}\right\} =0,
\end{align}
as usual. In addition, the modified completeness relations are given by \cite{Cabral:2025mjf}
\begin{eqnarray}
\sum_{s}u^{s}(p)\overline{u}^{s}(p)  =  \frac{(m-\mu\gamma_{5})\slashed{p}+M^{2}}{2M^{2}},\\
\sum_{s}\upsilon^{s}(p)\overline{\upsilon}^{s}(p)  =  \frac{(m-\mu\gamma_{5})\slashed{p}-M^{2}}{2M^{2}},
\end{eqnarray}
where the Feynman slash notation, $\slashed{p}=\gamma^\mu p_\mu$, has been used. 

It is customary to express the modified mass $M$ in terms of the actual mass $m$ by taking $\mu=\lambda m$, where $\lambda$  is a small correction due to the non-Hermiticity. This leads to the relation
\begin{equation}
M^2=m^2(1-\lambda^2).
\end{equation}
For electrons, it has been found that $\lambda_e=2.45\times10^{-5}$ \cite{Cabral:2025mjf}.

With these results we are now able to compute any non-Hermitian scattering process in the presence of a background thermal reservoir at tree level within the TFD formalism (in terms of the scattering matrix), which will be briefly introduced in the next subsection.

\subsection{The TFD formalism}\label{tfd}

To describe thermal scattering amplitudes, we have several formalisms that introduce a temperature parameter based on certain assumptions, including those using imaginary- or real-time transformations.
One of the simplest and powerful real-time formalism is the thermal field dynamics, which allows to evaluate thermal contributions in a scattering process in a similar fashion of calculations in the $T=0$ theory \cite{temp000, khannatfd}.

The TFD approach is characterized by doubling the Hilbert space $\mathbb{H}$ with a copy $\widetilde{\mathbb{H}}$,  referred to as the \textit{tilde} space, which acts as a heat bath in the field theory \cite{temp000, khannatfd}.
In this framework, particle dynamics take place in the thermal (or total) space $\mathbb{H}_\beta=\mathbb{H}\otimes\widetilde{\mathbb{H}}$, where $\beta=T^{-1}$ represents the temperature parameter. The thermal average of a quantum operator $\mathcal{O}$ is given by
\begin{eqnarray}
\mathcal{O}=\bra{0(\beta)}\mathcal{O}\ket{0(\beta)},
\end{eqnarray} 
where $\ket{0(\beta)}\in\mathbb{H}_\beta$ is the thermal vacuum.

The tilde operators $\widetilde{\mathcal{O}}$, defined on the doubled space $\widetilde{\mathbb{H}}$, act in a similar way as the operators $\mathcal{O}\in\mathbb{H}$.
These operators satisfy some conjugation rules, necessary for the formal development of the scattering matrix, which can be summarized as follows
\begin{eqnarray}
\widetilde{\mathcal{O}_1\mathcal{O}_2+c\mathcal{O}_3}=\widetilde{\mathcal{O}}_1\widetilde{\mathcal{O}}_2+c^{*}\widetilde{\mathcal{O}}_3,\quad\quad \widetilde{\mathcal{O}_4^\dagger+\widetilde{\mathcal{O}}_5}=\widetilde{\mathcal{O}}_4^\dagger-\zeta \mathcal{O}_5,
\end{eqnarray}
where $c$ is a complex number and $\zeta=1$ ($\zeta=-1$) for fermion (boson) variables.

Moreover, using the Bogoliubov transformation we can construct the creation and annihilation operators for fermionic particles in both spaces. Consequently, the thermal version of these operators can be expressed as
\begin{eqnarray}
a_p^s(\beta)=U(\beta,p_0)a_p^s-V(\beta,p_0)(\widetilde{a}_p^s)^\dagger,\quad\quad (a_p^s(\beta))^\dagger=U(\beta,p_0)(a_p^s)^\dagger-V(\beta,p_0)\widetilde{a}_p^s,\label{eq01}
\end{eqnarray}
where $s$ is the spin index. The weighting $\beta-$coefficients are thermal functions defined as
\begin{eqnarray}
U(\beta,p_o)=\sqrt{e^{\beta p_0}n_f(p_0)},\quad\quad V(\beta,p_o)=\sqrt{n_f(p_0)}
\end{eqnarray}
with $n_f(p_0)=(1+e^{\beta p_0})^{-1}$ being the Fermi-Dirac distribution.
The fermionic thermal operators satisfy the following anticommutation relation
\begin{eqnarray}
\{a_p^s(\beta),(a_q^r(\beta))^\dagger\}=N_p\delta_{sr}\delta^{ (3)}\left(\vec{p}-\vec{q}\right),
\end{eqnarray}
where $N_p$ is a normalization factor. The same discussion applies to the thermal antiparticle operators $b_p^s(\beta)$.

For bosons, the discussion is analogous. However, in this case, the Bogoliubov transformation takes the form
\begin{eqnarray}
d_p^\lambda(\beta)=U^\prime(\beta,p_0)d_p^\lambda+V^\prime(\beta,p_0)(\widetilde{d}_p^\lambda)^\dagger,\quad\quad (d_p^\lambda(\beta))^\dagger=U^\prime(\beta,p_0)(d_p^\lambda)^\dagger+V^\prime(\beta,p_0)\widetilde{d}_p^\lambda,
\end{eqnarray}
with the thermal functions being
\begin{eqnarray}
U^\prime(\beta,p_o)=\sqrt{e^{\beta p_0}n_b(p_0)},\quad\quad V^\prime(\beta,p_o)=\sqrt{n_b(p_0)},
\end{eqnarray}
where $n_b(p_0)=(e^{\beta p_0}-1)^{-1}$ is the Bose-Einstein distribution.

Furthermore, the thermal bosonic propagator in the position space is given by
\begin{eqnarray}
D_\beta^{\mu\nu}(x,y)=-i\int \frac{d^4q}{(2\pi)^4}e^{-iq(x-y)}\Delta_\beta^{\mu\nu}(q^2),
\end{eqnarray}
with $\Delta_\beta^{\mu\nu}(q^2)=\eta^{\mu\nu}\Delta_\beta(q^2)$ being the momentum space propagator\footnote{The following convention for the metric tensor has been used: $\text{diag}{(\eta_{\mu\nu})}=(1,-1,-1,-1)$.}, where
\begin{eqnarray}
\Delta_\beta(q^2)=\frac{1}{q^2}-\frac{2\pi i}{e^{\beta|q_0|}-1}\delta(q^2).
\end{eqnarray}

With the above discussion we have established all the necessary ingredients to determine a thermal non-Hermitian scattering process, in which we consider Bhabha scattering, specifically the reaction $e^+ e^-\to e^+ e^-$.
Our purpose is to compute the Bhabha process differential cross section in order to highlight its modification from the standard QED values, as well as make use of the precision data of this process to set stringent bounds upon the axial-vector coupling $g_a$.

\section{The Bhabha scattering process}
\label{scattering}

In order to develop the S-matrix contribution \footnote{Details about the construction of transition amplitudes within non-Hermitian models were discussed in \cite{Ohlsson:2019noy,Alexandre:2020gah}.} related with the Bhabha scattering process, which consists of $e^+e^- \to e^+e^-$, we must define the respective initial and final thermal states
\begin{eqnarray}
\ket{i(\beta)}&=&\sqrt{2E_{p_i}}\sqrt{2E_{k_i}}a^\dagger_\beta(p_i,s_1)b^\dagger_\beta(k_i,s_2)\ket{0(\beta)},\label{statei} \\ 
\ket{f(\beta)}&=&\sqrt{2E_{p_f}}\sqrt{2E_{k_f}}a^\dagger_\beta(p_f,s_3)b^\dagger_\beta(k_f,s_4)\ket{0(\beta)},\label{statef}
\end{eqnarray}
where the pre-factors $\sqrt{2E}$ ensure, as usual, proper normalization.

\begin{figure}[ht]
\includegraphics[scale=.5]{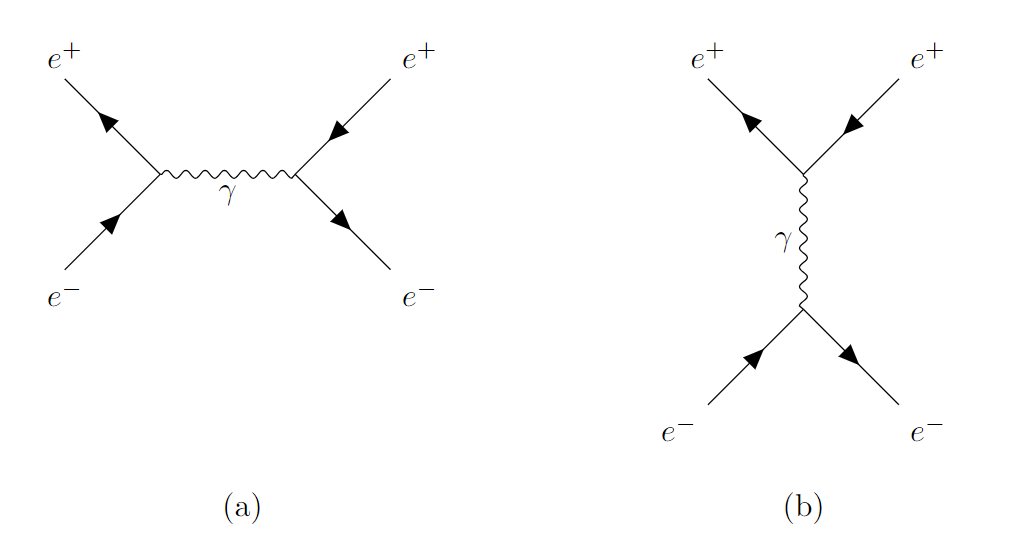}
\caption{Feynman diagrams for the Bhabha scattering process, illustrating the $s-$ channel (a) and  $t-$ channel (b) contributions.}
\label{fig1}
\end{figure}

In this way, the corresponding second-order scattering matrix element becomes
\begin{eqnarray}
S^{(2)}=\frac{(-i)^2}{2!}\int d^4xd^4y\mathcal{T}\left\{\mathcal{L}_I(x)\mathcal{L}_I(y)\right\},
\end{eqnarray}
where $\mathcal{T}$ denotes the time-ordering operator. This leads to the following expression for the transition amplitude
\begin{eqnarray}
\mathcal{M}=\bra{f(\beta)}S\ket{i(\beta)}=\mathcal{M}_a+\mathcal{M}_b,
\end{eqnarray}
with $\mathcal{M}_a$ and $\mathcal{M}_b$ being the probability amplitudes describing by the $s-$ and $t-$ channel Feynman diagrams depicted in Figure \ref{fig1}.

Hence, taking into account the respective Feynman rules and the non-hermitian vertex $-i\gamma_\mu\left(g_v+g_a\gamma^5\right)$, the diagram (a) reads
\begin{eqnarray}
\mathcal{M}_a&=&-4M^2F(\beta)\bar{v}^{s_2}(k_i)\gamma_\mu\left(g_v+g_a\gamma^5\right) u^{s_1}(p_i)\Delta_\beta^{\mu\nu}(s)\bar{u}^{s_3}(p_f)\gamma_\nu\left(g_v+g_a\gamma^5\right) v^{s_4}(k_f).
\end{eqnarray}
Analogously, for the second diagram we have
\begin{eqnarray}
\mathcal{M}_b=4M^2F(\beta)\bar{v}^{s_2}(k_i)\gamma_\mu\left(g_v+g_a\gamma^5\right) v^{s_4}(k_f)\Delta_\beta^{\mu\nu}(t)\bar{u}^{s_3}(p_f)\gamma_\nu\left(g_v+g_a\gamma^5\right) u^{s_1}(p_i)
\end{eqnarray}
where $s=E_\text{{cm}}^2$ and $t=2(M^2-E^2)(1-\cos{\theta})$ are the Mandelstam variables in the centre-of-mass (COM) frame. The thermal function $F(\beta)$, which appears as an overall multiplication factor of the amplitudes, is given by
\begin{eqnarray}
F(\beta)=\frac{1}{4}\left[1+\tanh(\frac{\beta\sqrt{s}}{2})\right]^2. \label{eq04}
\end{eqnarray} 
Actually, this function is obtained from the thermal factors $U(\beta,p_0)$ coming from the fermionic creation and annihilation operators, after performing the respective Bogoliubov's transformations \eqref{eq01} into the states \eqref{statei} and \eqref{statef}.

Moreover, since the thermal differential cross section for the Bhabha scattering is defined as
\begin{eqnarray}
\left(\frac{d\sigma}{d\Omega}\right)_{\text{Bhabha}}&=&\frac{|\vec{p}_i||\vec{p}_f|}{(4\pi s)^2}\langle|\mathcal{M}|^2\rangle ,\label{eq03a}
\end{eqnarray}
we can consider the average over the spin states, and thus we obtain the squared transition amplitude
\begin{eqnarray} \label{eq100}
\langle|\mathcal{M}|^2\rangle&=&\langle|\mathcal{M}_a|^2\rangle+\langle|\mathcal{M}_b|^2\rangle+2\Re{\langle\mathcal{M}_a^\dagger\mathcal{M}_b\rangle}.
\end{eqnarray}
Hence, evaluating explicitly the squared amplitudes in \eqref{eq100}, we cast the differential cross-section in the form
\begin{eqnarray}
\left(\frac{d\sigma}{d\Omega}\right)_{\text{Bhabha}}= F^2(\beta)\biggl(\frac{d\sigma}{d\Omega}\biggr)_\beta,\label{eq03}
\end{eqnarray}
in which we have defined
\begin{align}\label{eq07}
\biggl(\frac{d\sigma}{d\Omega}\biggr)_\beta &=\frac{|\Delta_\beta(s)|^2}{16\pi^2s^2}\left[g_a^4\chi_1+g_a^3g_v\chi_2+g_a^2g_v^2\chi_3+g_ag_v^3\chi_4+g_v^4\chi_5\right] \cr
&+\frac{|\Delta_\beta(t)|^2}{16\pi^2s^2}\left[g_a^4\Gamma_1+g_a^3g_v\Gamma_2+g_a^2g_v^2\Gamma_3+g_ag_v^3\Gamma_4+g_v^4\Gamma_5\right]
\cr
&+\frac{\Re{\Delta_\beta(s)\Delta^{*}_\beta(t)}}{16\pi^2s^2}\left[g_a^4\xi_1+g_a^3g_v\xi_2+g_a^2g_v^2\xi_3+g_ag_v^3\xi_4+g_v^4\xi_5\right],
\end{align}
by simplicity of notation, the auxiliary functions $\chi_i$, $\Gamma_i$ and $\xi_i$, with $i=1,...,5$,  are presented in Appendix \ref{app}.

Since we wish to set stringent bounds upon the free parameters of the non-hermitian QED model it is interesting to consider the leading contributions obtained from the high-energy limit $m=0$ (but keeping non-hermitian effects in terms of $\lambda \neq 0$).
Hence, under this consideration and some simplification, we can write
\begin{align}
\left(\frac{d\sigma}{d\Omega}\right)_\beta^{m\to0}&=\frac{(g_a^4+g_v^4)}{512\pi^2s}\left[4\lambda^4\frac{ 56 \cos \theta +28 \cos (2 \theta )+8 \cos (3 \theta )+\cos (4 \theta
   )}{ (\cos \theta -1)^2}\right. \cr
      &+\left.\frac{28 \cos (2 \theta )+\cos (4 \theta )+140 \lambda ^2+99}{ (\cos \theta -1)^2}\right] \cr
   &+\frac{g_a^2g_v^2}{256\pi^2 s}
   \left[\frac{16 \left(6 \lambda ^2+1\right) \cos (3 \theta )+\left(12 \lambda ^2+1\right) \cos (4
   \theta )+420 \lambda ^2-29}{(\cos \theta -1)^2}\right.\nonumber\\
   &+\left.     \frac{112 \left(6 \lambda ^2+1\right) \cos \theta +28 \left(12 \lambda ^2+1\right) \cos    (2 \theta )}{(\cos \theta -1)^2}    \right]
   -\lambda\frac{(g_ag_v^3+g_vg_a^3)}{512\pi^2 s}  \sin ^8\theta  \csc ^{12}\left(\frac{\theta }{2}\right)\nonumber\\
   &+\frac{(g_a^4+g_v^4)s}{8(e^{\beta\sqrt{s}/2}-1)^2}\left[\left(8 \lambda ^2+1\right)  \cos ^4\left(\frac{\theta }{2}\right)+1\right]\delta^2\left[\frac{s}{2}(\cos{\theta}-1)\right]\nonumber\\
   &+\frac{3g_a^2g_v^2s}{4(e^{\beta\sqrt{s}/2}-1)^2}\left[\left(8 \lambda ^2+1\right)  \cos ^4\left(\frac{\theta }{2}\right)-\frac{1}{3}\right]\delta^2\left[\frac{s}{2}(\cos{\theta}-1)\right]\nonumber\\
   &-\frac{2\lambda(g_ag_v^3+g_a^3g_v)s}{(e^{\beta\sqrt{s}/2}-1)^2}  \cos ^4\left(\frac{\theta }{2}\right)\delta^2\left[\frac{s}{2}(\cos{\theta}-1)\right].\label{eq02}
\end{align}
It is worth noting that squared delta functions evaluated at the same point do not pose a problem here because they are a common feature present in all calculations using the TFD formalism \cite{regdelta}.
Moreover, it should be emphasized that the expression \eqref{eq02} is valid in the unbroken PT-symmetry regime $\lambda \ll 1$.

 We now consider some interesting limits. The first is the high-temperature limit ($\beta \to 0$), in which the cross section becomes
\begin{eqnarray}
\left(\frac{d\sigma}{d\Omega}\right)^{m\to0}_{\beta\to0}&=&\frac{(g_a^4+g_v^4)}{512\pi^2s}\left[4\lambda^4\frac{ 56 \cos \theta +28 \cos (2 \theta )+8 \cos (3 \theta )+\cos (4 \theta
   )}{ (\cos \theta -1)^2}\right. \cr
      &+&\left.\frac{28 \cos (2 \theta )+\cos (4 \theta )+140 \lambda ^2+99}{ (\cos \theta -1)^2}\right] \cr
   &+&\frac{g_a^2g_v^2}{256\pi^2 s}
   \left[\frac{16 \left(6 \lambda ^2+1\right) \cos (3 \theta )+\left(12 \lambda ^2+1\right) \cos (4
   \theta )+420 \lambda ^2-29}{(\cos \theta -1)^2}\right.\nonumber\\
   &+&\left.     \frac{112 \left(6 \lambda ^2+1\right) \cos \theta +28 \left(12 \lambda ^2+1\right) \cos    (2 \theta )}{(\cos \theta -1)^2}    \right]
   -\lambda\frac{(g_ag_v^3+g_vg_a^3)}{512\pi^2 s}  \sin ^8\theta  \csc ^{12}\left(\frac{\theta }{2}\right)\nonumber\\&+&\frac{g_a^4+g_v^4}{2\beta^2}\left[\left(8 \lambda ^2+1\right)  \cos ^4\left(\frac{\theta }{2}\right)+1\right]\delta^2\left[\frac{s}{2}(\cos{\theta}-1)\right]\nonumber\\
   &+&\frac{3g_a^2g_v^2}{\beta^2}\left[\left(8 \lambda ^2+1\right)  \cos ^4\left(\frac{\theta }{2}\right)-\frac{1}{3}\right]\delta^2\left[\frac{s}{2}(\cos{\theta}-1)\right]\nonumber\\
   &-&\frac{8\lambda(g_ag_v^3+g_a^3g_v)}{\beta^2}  \cos ^4\left(\frac{\theta }{2}\right)\delta^2\left[\frac{s}{2}(\cos{\theta}-1)\right].
\end{eqnarray} 
In addition, from Eq. \eqref{eq04}, it is obtained that $F(\beta) \to 1/4$, and the full differential cross section, following Eq. \eqref{eq02}, under this condition, becomes
\begin{eqnarray}
\left(\frac{d\sigma}{d\Omega}\right)_\text{Bhabha}&=&\frac{1}{16}\left(\frac{d\sigma}{d\Omega}\right)^{m\to0}_{\beta\to0},
\end{eqnarray}
implying that the reaction increases proportionally to $T^2$ and that these thermal terms have a distinct dependence on the squared energy $s$ (appearing only on the delta functions, playing the role of filters for such values).
This behavior shows an increasing number of scattering events in hotter scenarios,   being very interesting and implies that quantities beyond the Standard Model (such as $\mu$ and $g_a$ in the present work) can be more easily detected in such cases.

Furthermore, in order to establish the comparison between the modified Bhabha differential cross section \eqref{eq03} and \eqref{eq02} with the experimental data it is necessary to take the zero temperature limit $\beta\to\infty$.
Thus,  we observe from the definition \eqref{eq04} that $F(\beta)\to1$, and also that the delta contributions in the Eq. \eqref{eq02} vanish.
Hence, in this regime the differential cross section behaves as
\begin{eqnarray} \label{eq110}
\left(\frac{d\sigma}{d\Omega}\right)_{\beta\to\infty}^{m\to0}&=&\frac{(\alpha_a^2+\alpha^2)}{32s}\left[4\lambda^4\frac{ 56 \cos \theta +28 \cos (2 \theta )+8 \cos (3 \theta )+\cos (4 \theta
   )}{ (\cos \theta -1)^2}\right.\nonumber\\
   &+&\left.\frac{28 \cos (2 \theta )+\cos (4 \theta )+140 \lambda ^2+99}{ (\cos \theta -1)^2}\right] \cr 
   &+&\frac{\alpha_a\alpha}{16 s}\left[\frac{112 \left(6 \lambda ^2+1\right) \cos  \theta}{(\cos \theta -1)^2}\right. \cr
   &+&\left.\frac{28 \left(12 \lambda ^2+1\right) \cos
   (2 \theta )+16 \left(6 \lambda ^2+1\right) \cos (3 \theta )+\left(12 \lambda ^2+1\right) \cos (4
   \theta )+420 \lambda ^2-29}{(\cos \theta -1)^2}\right]\nonumber\\
   &-&\frac{\lambda \sqrt{\alpha \alpha _a}\left(\alpha +\alpha_a   \right)}{32 s}  \sin ^8\theta  \csc ^{12}\left(\frac{\theta }{2}\right),
\end{eqnarray}
where $\alpha_a=g_a^2/4\pi$ and $\alpha=g_v^2/4\pi$ are the axial and QED fine structure constants, respectively.
It is interesting to observe that the energy profile in the result \eqref{eq110} is simply $1/s$, this is the same behavior of the QED expression.

Finally, from the expression \eqref{eq110} we can determine the non-Hermitian deviation in relation to the tree-level QED results and use to estimate a bound for the $\alpha_a$ parameter.
Hence, we can match these non-Hermitian contributions with the associated error for the Bhabha differential cross-section,
\begin{equation} \label{eq05}
\delta \left(\frac{d\sigma}{d\Omega}\right)_\text{non-H}= \frac{|\left(\frac{d\sigma}{d\Omega}\right)_\text{MEAS}-\left(\frac{d\sigma}{d\Omega}\right)_\text{QED}|}{\left(\frac{d\sigma}{d\Omega}\right)_\text{QED}}
\end{equation}
in which $\left(\frac{d\sigma}{d\Omega}\right)_\text{Meas}$ corresponds to the experimental value of the Bhabha scattering, while the QED expression is
\begin{eqnarray}
\left(\frac{d\sigma}{d\Omega}\right)_\text{QED}=\frac{\alpha^2}{4s}\frac{(\cos^2{\theta}+3)^2}{(\cos{\theta}-1)^2}.\label{eq06}
\end{eqnarray}

The data for the QED and measured differential cross-sections at given angular distribution are presented in Table \ref{tab1}.
There, we also establish a series of bounds upon the coupling $\alpha_a$ obtained by solving \eqref{eq05}.
\begin{table}[h!]
\centering
\begin{minipage}{0.48\textwidth}
\centering
\begin{tabular}{|c|c|c|c|}
\hline
$\cos{\theta}$ & $s\left(\frac{d\sigma}{d\Omega}\right)_\text{Meas}$  & $s\left(\frac{d\sigma}{d\Omega}\right)_\text{QED}$ & $\alpha_a(10^{-4})$ \\ \hline
$-0.525$ & $22.8$  & $23.9$  & $1.76$ \\ \hline
$-0.475$ & $24.9$  & $24.8$  & $0$ \\ \hline
$-0.425$ & $24.8$  & $25.8$  & $1.51$ \\ \hline
$-0.375$ & $24.7$  & $27.0$  & $3.36$ \\ \hline
$-0.325$ & $26.8$  & $28.5$  & $2.30$ \\ \hline
$-0.275$ & $29.4$  & $30.2$  & $1.01$ \\ \hline
$-0.225$ & $31.5$  & $32.2$  & $0.82$ \\ \hline
$-0.175$ & $33.3$  & $34.5$  & $1.43$ \\ \hline
$-0.125$ & $36.0$  & $37.3$  & $1.44$ \\ \hline
$-0.075$ & $38.7$  & $40.5$  & $2.01$ \\ \hline
$-0.025$ & $43.9$  & $44.4$  & $0.56$ \\ \hline
\end{tabular}
\end{minipage}
\hfill
\begin{minipage}{0.48\textwidth}
\centering
\begin{tabular}{|c|c|c|c|}
\hline
$\cos{\theta}$ & $s\left(\frac{d\sigma}{d\Omega}\right)_\text{Meas}$  & $s\left(\frac{d\sigma}{d\Omega}\right)_\text{QED}$ & $\alpha_a(10^{-4})$  \\ \hline
$0.025$  & $47.8$  & $49.1$  & $1.32$ \\ \hline
$0.075$  & $54.3$  & $54.7$  & $0.44$ \\ \hline
$0.125$  & $62.3$  & $61.6$  & $0$ \\ \hline
$0.175$  & $69.1$  & $69.99$ & $0.799$ \\ \hline
$0.225$  & $78.4$  & $80.4$  & $1.77$ \\ \hline
$0.275$  & $91.6$  & $93.3$  & $1.58$ \\ \hline
$0.325$  & $111.7$ & $109.8$ & $0$ \\ \hline
$0.375$  & $128.4$ & $130.96$ & $2.78$ \\ \hline
$0.425$  & $156.5$ & $158.7$  & $3.08$ \\ \hline
$0.475$  & $194.6$ & $195.8$  & $3.24$ \\ \hline
$0.525$  & $250.8$ & $246.6$  & $8.9$ \\ \hline
\end{tabular}
\end{minipage}

\caption{The measured and QED differential cross-sections (in nb GeV$^{2}$) in terms of the the angular distribution and its respective limits upon the $\alpha_a$ parameter (estimated using \ref{eq05}). The experimental points were obtained by \cite{29gev} at energy of $\sqrt{s}=29$ GeV. }
\label{tab1}
\end{table}

\begin{figure}[ht]
\includegraphics[scale=0.8]{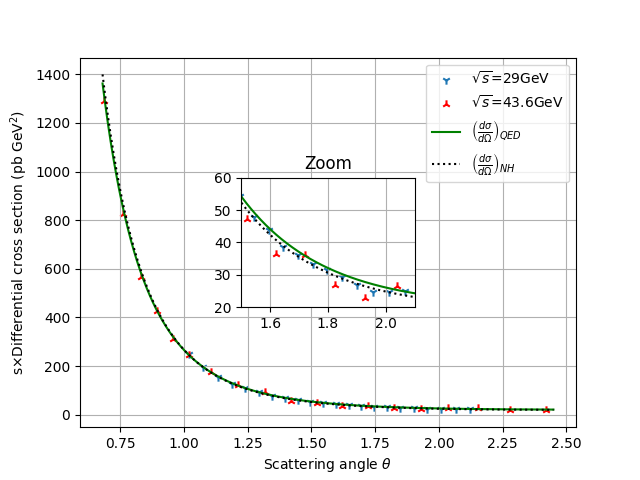}
\caption{The differential cross sections (multiplied by the squared centre of mass energy) for the QED  (solid green line) and non-hermitian QED (dashed black line) in terms of the experimental data of \cite{29gev} at $\sqrt{s}=29$ GeV (down blue marker) and \cite{43gev} at $\sqrt{s}=43.6$ GeV (up red marker) in terms of $\cos{\theta}$.}\label{fig2}
\end{figure}

Moreover, one can take the mean value of the coupling $\alpha_a$ values on the \ref{tab1} and obtain that $\alpha_a = 1.82 \times 10^{-4}$, which implies $\alpha_a \approx 1/5494$, which is significantly smaller than the fine structure constant $\alpha_a \ll \alpha \simeq 1/137$ and is consistent with our idea to consider the non-hermitian effects as small perturbations. 
In order to highlight the aspects of these findings, we present a plot to compare the QED and zero-temperature non-Hermitian results with the experimental values from two different experiments, these are depicted in Figure \ref{fig2}. The values $g_a = 0.0478$ and $\lambda = 2.45 \times 10^{-5}$ have been used.

As one would expect, due to the smallness of the non-hermitian parameters, the non-Hermitian QED and usual QED exhibit very similar curve profile in most regions with respect to the scattering angle. In certain regions, such as $1.5 \leq \theta \leq 2.1$ (zoomed in Figure~\ref{fig2}), one observes that the non-Hermitian curve presents a better agreement with the experimental data for both data presented.

\section{Conclusions}
\label{secconclusion}

This paper investigates Bhabha scattering ($e^{+}e^{-} \to e^{+}e^{-}$) under the influence of non-Hermiticity and temperature effects.
For the first modification, we develop a framework for the non-Hermitian QED by introducing an axial mass term $\mu$ and a vector-axial coupling current with the gauge field, which yields modification into the Feynman rules, the fermion propagator and an additional vertex, $-i\gamma_\mu(g_v + g_a\gamma^5)$.
To account for temperature effects, we employ the thermofield dynamics formalism, which incorporates thermal corrections by doubling the respective Fock space (in terms of Bogoliubov transformation).
In this framework, thermal fermionic (bosonic) operators are expressed as linear combinations of their zero-temperature counterparts, weighted by the Fermi-Dirac (Bose-Einstein) distribution.

Using the TFD approach we computed the thermal differential cross section for the Bhabha scattering in the non-Hermitian QED, because its high precision data allow us to set stringent bounds upon the free parameters.
We revised the main aspects about the free field solutions for this non-Hermitian model and also how the temperature effects are incorporated into the scattering amplitudes.
Since the respective expression for the thermal differential cross-section is very extensive, it was necessary to consider some approximation in order to select the leading contributions from non-hermiticity: this was achieved by treating the non-expansion-based parameter $\lambda \ll 1$ as a small perturbation.

The behavior of scattering at high temperature was explored, which exhibits a $T^2$ profile, increasing the number of scattering events in hotter scenarios (these thermal terms have a distinct dependence on the squared energy $s$ appearing only on the delta functions, playing the role of filters for such values). This result is very interesting and implies that quantities beyond the Standard Model (such as $\mu$ and $g_a$ in the present work) can be more easily detected in such cases. Furthermore, these quantities could play a significant role in the scattering process, dictating how the particles behave in reality. However, due to the nature of the phenomenological values, which are measured under conditions that simulate the $\beta \to \infty$ scenario, the high-temperature expression is not used to obtain constraints on the non-Hermitian parameters of the theory. 

On the other hand, the high-energy and zero-temperature limits of this quantity were also considered to examine the impact of non-Hermicity on the Bhabha scattering experimental data.
A comparison yielded a bound value for the axial coupling constant, $\alpha_a \approx 1/5494$ (satisfying $\alpha_a \ll \alpha$), for the reaction involving electrons and positrons.
A plot summarizing these findings is shown in Figure \ref{fig2} and shows good agreement between the zero-temperature non-hermitian QED predictions and the experimental data, particularly in the region $1.5 \leq \theta \leq 2.1$.

{A last remark we would like to present is about the underlying physical principles in regard to the Lagrangian density \eqref{lagran}. This given model was developed in terms of PT-symmetry, as discussed in \cite{Alexandre:2015kra}.
However, though the PT-symmetry and the Pseudo-Hermiticity coexist in many cases, they belong to different types of discrete symmetries \cite{Mostafazadeh:2008pw} and  one should be careful when constructing a QFT model.
On the other hand, it is worth to mention that in the case when both coexist they exactly led to the same physical content although some formal aspects differ.
An interesting example is the study of the behavior of the Higgs mechanism and Goldstone theorem which were discussed in the PT-symmetric case and pseudo-Hermitian approach leading to the same physical masses of the vector gauge bosons, but to contrasting behavior of some formal aspects at the exceptional points \cite{Alexandre:2018uol,Mannheim:2018dur,Alexandre:2019jdb,Fring:2019xgw,Fring:2020bvr}.}

\section*{Acknowledgments}

This work by A. F. S. is partially supported by National Council for Scientific and Technological
Development - CNPq project No. 312406/2023-1. D. S. C. thanks CAPES for financial support.
R.B. acknowledges partial support from Conselho
Nacional de Desenvolvimento Cient\'ifico e Tecnol\'ogico (CNPq Project No.~ 306769/2022-0).

\appendix 
\section{Auxiliary functions for Eq. \eqref{eq07}}
\label{app}

For the sake of notation, the differential cross section \eqref{eq07} has been presented in terms of new functions defined as
\begin{eqnarray}
\chi_1&=&\frac{s^3}{8}\mathcal{G}(\Xi^{16,0},\Xi^{4,1},3\Xi^{4,1})-\frac{(sm)^2}{2}\mathcal{G}(\Xi^{32,0},\Xi^{9,3},\Xi^{19,9}),\nonumber\\
\chi_2&=&-8 \lambda  s^3 \cos ^4\left(\frac{\theta }{2}\right)+4 \lambda  (sm)^2\mathcal{G}(8,3,7),\nonumber\\
\chi_3&=&\frac{s^3}{4}\mathcal{G}(8\Xi^{6,1},\Xi^{12,1},3\Xi^{12,1})-{(sm)^2}{}\mathcal{G}(16\Xi^{5,1},3\Xi^{11,1},5\Xi^{11,1}),\nonumber\\
\chi_4&=&-8 \lambda  s^3 \cos ^4\left(\frac{\theta }{2}\right)+4 \lambda  (sm)^2 \mathcal{G}(8,3,3),\nonumber\\
\chi_5&=&\frac{s^3}{8}\mathcal{G}(16\Xi^{1,0},\Xi^{4,1},3\Xi^{4,1})-\frac{(sm)^2}{2}\mathcal{G}(32\Xi^{1,0},3\Xi^{3,1},\Xi^{11,1}),
\end{eqnarray}
for the $s-$channel, in which we defined the following functions,
\begin{eqnarray}
\mathcal{G}(a,b,c)=a\cos\theta+b\cos(2\theta)+c,\quad\quad\Xi^{a,b}=a\lambda^2+b.
\end{eqnarray}
Moreover,
\begin{eqnarray}
\Gamma_1&=&\frac{s^3}{16}\mathcal{G}(4\Xi^{8,1},\Xi^{8,1},\Xi^{24,11})-\frac{(sm)^2}{4}\mathcal{G}(12\Xi^{5,1},3\Xi^{7,1},\Xi^{15,17})\nonumber\\
\Gamma_2&=&-8 \lambda  s^3 \cos ^4\left(\frac{\theta }{2}\right)+4 \lambda  (sm)^2  \mathcal{G}(9,3,4)\nonumber\\\Gamma_3&=&\frac{s^3}{8}\mathcal{G}(12\Xi^{8,1},3\Xi^{8,1},\Xi^{72,1})-{(sm)^2}{}\mathcal{G}(24\Xi^{7,1},9\Xi^{7,1},\Xi^{121,-1})\nonumber\\
\Gamma_4&=&-8 \lambda  s^3 \cos ^4\left(\frac{\theta }{2}\right)+4 \lambda  (sm)^2 \mathcal{G}(7,3,6)\nonumber\\
   \Gamma_5&=&\frac{s^3}{16}\mathcal{G}(4\Xi^{8,1},\Xi^{8,1},\Xi^{24,11})-\frac{(sm)^2}{4}\mathcal{G}(4\Xi^{13,1},3\Xi^{7,1},\Xi^{23,25})
\end{eqnarray}
for the $t-$ channel. Finally,
\begin{eqnarray}
   \xi_1&=& s^3\Xi^{8,1} \cos ^4\left(\frac{\theta }{2}\right)-2 (sm)^2 \cos ^2\left(\frac{\theta }{2}\right) \mathcal{G}(3\Xi^{7,1},0,\Xi^{9,1}),\nonumber\\
\xi_2&=&-16 \lambda  s^3  \cos ^4\left(\frac{\theta }{2}\right)+4 \lambda  (sm)^2 \mathcal{G}(17,6,11),\nonumber\\
   \xi_3&=&6 \Xi^{8,1} s^3 \cos ^4\left(\frac{\theta }{2}\right)-4 (sm)^2 \cos ^2\left(\frac{\theta }{2}\right)\mathcal{G}(9\Xi^{7,1},0,\Xi^{20,4}),\nonumber\\\xi_4&=&-16 \lambda  s^3 \cos ^4\left(\frac{\theta }{2}\right)+12 \lambda  (sm)^2 \mathcal{G}(5,2,3),\nonumber\\\xi_5&=&s^3\Xi^{8,1} \cos ^4\left(\frac{\theta }{2}\right)-2 (sm)^2 \cos ^2\left(\frac{\theta }{2}\right) \mathcal{G}(3\Xi^{7,1},0,\Xi^{7,-1}),
\end{eqnarray}
for the mixed channel.
It worth to mention that in order to keep the leading non-hermitian contributions, it has been considered the calculations up to second order on the parameters $m$ and $\lambda$. 


\global\long\def\link#1#2{\href{http://eudml.org/#1}{#2}}
 \global\long\def\doi#1#2{\href{http://dx.doi.org/#1}{#2}}
 \global\long\def\arXiv#1#2{\href{http://arxiv.org/abs/#1}{arXiv:#1 [#2]}}
 \global\long\def\arXivOld#1{\href{http://arxiv.org/abs/#1}{arXiv:#1}}



\begin{thebibliography}{99}





\bibitem{peskin} M. E. Peskin, ``An introduction to quantum field theory''. CRC press, (2018).

\bibitem{greiner} W. Greiner and J. Reinhardt, ``Quantum electrodynamics''. Springer Science \& Business Media, (2008).

\bibitem{Charneski:2012py}
B.~Charneski, M.~Gomes, R.~V.~Maluf and A.~J.~da Silva,
``Lorentz violation bounds on Bhabha scattering,''
\doi{10.1103/PhysRevD.86.045003}{Phys. Rev. D \textbf{86} (2012), 045003}
\arXiv{1204.0755}{hep-ph}.

\bibitem{Casana:2012vu}
R.~Casana, M.~M.~Ferreira, R.~V.~Maluf and F.~E.~P.~dos Santos,
``Effects of a CPT-even and Lorentz-violating nonminimal coupling on the electron-positron scattering,''
\doi{10.1103/PhysRevD.86.125033}{Phys. Rev. D \textbf{86} (2012), 125033}
\arXiv{1212.6230}{hep-th}.

\bibitem{Bufalo:2015eia}
R.~Bufalo,
``On the Bhabha scattering for z = 2 Lifshitz QED,''
\doi{10.1142/S0217751X15500864}{Int. J. Mod. Phys. A \textbf{30} (2015) no.16, 1550086}
\arXiv{1505.03974}{hep-th}. 


\bibitem{deBrito:2016zav}
G.~P.~de Brito, J.~T.~Guaitolini, Junior, D.~Kroff, P.~C.~Malta and C.~Marques,
``Lorentz violation in simple QED processes,''
\doi{10.1103/PhysRevD.94.056005}{Phys. Rev. D \textbf{94} (2016) no.5, 056005}
\arXiv{1605.08059}{hep-ph}.


\bibitem{Bender:1998ke}
C.~M.~Bender and S.~Boettcher,
``Real spectra in non-Hermitian Hamiltonians having PT symmetry,''
\doi{10.1103/PhysRevLett.80.5243} {Phys. Rev. Lett. \textbf{80} (1998), 5243-5246}
\arXivOld{physics/9712001}.

\bibitem{Mostafazadeh:2001jk}
A.~Mostafazadeh,
``PseudoHermiticity versus PT symmetry. The necessary condition for the reality of the spectrum,''
\doi{10.1063/1.1418246} {J. Math. Phys. \textbf{43} (2002), 205-214}
\arXivOld{math-ph/0107001}.

\bibitem{Bender:2005tb} 
C.~M.~Bender,
``Introduction to PT-Symmetric Quantum Theory,''
\doi{10.1080/00107500072632} {Contemp. Phys. \textbf{46} (2005), 277-292}
\arXivOld{quant-ph/0501052}.

\bibitem{Bender:2019}
C.~M.~Bender et al,
`` PT Symmetry in Quantum and Classical Physics''
WORLD SCIENTIFIC (EUROPE), 2019

\bibitem{Alexandre:2023afi}
J.~Alexandre, M.~Dale, J.~Ellis, R.~Mason and P.~Millington,
``Oscillation probabilities for a $\mathcal{PT}$-symmetric non-Hermitian two-state system,''
\arXiv{2302.11666}{quant-ph}. 

\bibitem{Simon:2018zrj}
M.~A.~Sim{\'o}n, A.~Buend{\'\i}a, A.~Kiely, A.~Mostafazadeh and J.~G.~Muga,
``$S$-matrix pole symmetries for non-Hermitian scattering Hamiltonians,''
\doi{10.1103/PhysRevA.99.052110}{Phys. Rev. A \textbf{99} (2019) no.5, 052110}
\arXiv{1811.06270}{quant-ph}.

\bibitem{novitsky}
A.~Novitsky,  D.~Lyakhov, D.~Michels, Alexander A.~Pavlov, Alexander S.~Shalin, and  Denis V. Novitsky,
``Unambiguous scattering matrix for non-Hermitian systems,''
\doi{10.1103/PhysRevA.101.043834}{Phys. Rev. A \textbf{101} (2020) no.4, 043834}

\bibitem{Pekduran:2025ljx}
B.~Pekduran, M.~Sar{\i}saman and E.~Ayd{\i}ner,
``Non-Hermitian gravitational wave scattering,''
\doi{10.1016/j.aop.2025.170109}{Annals Phys. \textbf{480} (2025), 170109}
\arXiv{2506.08567}{gr-qc}.


\bibitem{Alexandre:2022uns}
J.~Alexandre, J.~Ellis and P.~Millington,
``Discrete spacetime symmetries, second quantization, and inner products in a non-Hermitian Dirac fermionic field theory,''
\doi{10.1103/PhysRevD.106.065003}{Phys. Rev. D \textbf{106} (2022) no.6, 065003}
\arXiv{2201.11061}{hep-th}.

\bibitem{Sablevice:2023odu}
E.~Sablevice and P.~Millington,
``Poincar{\'e} symmetries and representations in pseudo-Hermitian quantum field theory,''
\doi{10.1103/PhysRevD.109.065012}{Phys. Rev. D \textbf{109} (2024) no.6, 6}
\arXiv{2307.16805}{hep-th}.


\bibitem{Li:2022prn}
W.~Li,
``Null bootstrap for non-Hermitian Hamiltonians,''
\doi{10.1103/PhysRevD.106.125021}{Phys. Rev. D \textbf{106} (2022) no.12, 125021}
\arXiv{2202.04334}{hep-th}.

\bibitem{Khan:2024mhc}
S.~Khan and H.~Rathod,
``Bootstrapping non-Hermitian quantum systems,''
\doi{10.1103/PhysRevD.111.105005}{Phys. Rev. D \textbf{111} (2025) no.10, 105005}
\arXiv{2409.06784}{hep-th}.

\bibitem{Chernodub:2024lkr}
M.~N.~Chernodub and P.~Millington,
``Anomalous dispersion, superluminality, and instabilities in two-flavor theories with local non-Hermitian mass mixing,''
\doi{10.1103/PhysRevD.109.105006}{Phys. Rev. D \textbf{109} (2024) no.10, 105006}
\arXiv{2401.06097}{hep-th}.


\bibitem{Alexandre:2015kra}
J.~Alexandre, C.~M.~Bender and P.~Millington,
``Non-Hermitian extension of gauge theories and implications for neutrino physics,''
\doi{10.1007/JHEP11(2015)111} {JHEP \textbf{11} (2015), 111}
\arXiv{1509.01203}{hep-th}. 


\bibitem{Alexandre:2020bet}
J.~Alexandre and N.~E.~Mavromatos,
``On the consistency of a non-Hermitian Yukawa interaction,''
\doi{10.1016/j.physletb.2020.135562}{Phys. Lett. B \textbf{807} (2020), 135562}
\arXiv{2004.03699}{hep-ph}.


\bibitem{Mavromatos:2022heh}
N.~E.~Mavromatos, S.~Sarkar and A.~Soto,
``Schwinger-Dyson equations and mass generation for an axion theory with a PT symmetric Yukawa fermion interaction,''
\doi{10.1016/j.nuclphysb.2022.116048}{Nucl. Phys. B \textbf{986} (2023), 116048}
\arXiv{2208.12436}{hep-ph}.

\bibitem{Fring:2019xgw}
A.~Fring and T.~Taira,
``Pseudo-Hermitian approach to Goldstone{\textquoteright}s theorem in non-Abelian non-Hermitian quantum field theories,''
\doi{10.1103/PhysRevD.101.045014}{Phys. Rev. D \textbf{101} (2020) no.4, 045014}
\arXiv{1911.01405}{hep-th}.

\bibitem{Fring:2020bvr}
A.~Fring and T.~Taira,
``Massive gauge particles versus Goldstone bosons in non-Hermitian non-Abelian gauge theory,''
\doi{10.1140/epjp/s13360-022-02889-z}{Eur. Phys. J. Plus \textbf{137} (2022) no.6, 716}
\arXiv{2004.00723}{hep-th}.


\bibitem{Felski:2022dsx}
A.~Felski, A.~Beygi and S.~P.~Klevansky,
``Thermodynamic properties of non-Hermitian Nambu{\textendash}Jona-Lasinio models,''
\doi{10.1103/PhysRevD.107.016015}{Phys. Rev. D \textbf{107} (2023) no.1, 016015}
\arXiv{2210.15503}{hep-ph}.

\bibitem{Mintz:2024soo}
B.~W.~Mintz, I.~Y.~Pinheiro and R.~Aquino,
``Oscillators with imaginary coupling: Spectral functions in quantum mechanics and quantum field theory,''
\doi{10.1103/PhysRevD.111.065014}{Phys. Rev. D \textbf{111} (2025) no.6, 065014}
\arXiv{2412.14064}{quant-ph}.


\bibitem{Arean:2019pom}
D.~Are{\'a}n, K.~Landsteiner and I.~Salazar Landea,
``Non-hermitian holography,''
\doi{10.21468/SciPostPhys.9.3.032}{SciPost Phys. \textbf{9} (2020) no.3, 032}
\arXiv{1912.06647}{hep-th}.


\bibitem{Arean:2024lzz}
D.~Arean and D.~Garcia-Fari{\~n}a,
``Holographic non-Hermitian lattices and junctions and their RG flows,''
\doi{10.1007/JHEP07(2025)276}{JHEP \textbf{07} (2025), 276}
\arXiv{2410.13584}{hep-th}.

\bibitem{Arean:2024gks}
D.~Arean, D.~Garcia-Fari{\~n}a and K.~Landsteiner,
``Strongly Coupled PT-Symmetric Models in Holography,''
\doi{10.3390/e27010013}{Entropy \textbf{27} (2025) no.1, 13}
\arXiv{2411.18471}{hep-th}.

\bibitem{Bender:2006fe}
C.~M.~Bender, D.~C.~Brody, H.~F.~Jones and B.~K.~Meister,
``Faster than Hermitian quantum mechanics,''
\doi{10.1103/PhysRevLett.98.040403} {Phys. Rev. Lett. \textbf{98} (2007), 040403}
\arXivOld{quant-ph/0609032}.
 
\bibitem{Makris_2008}
K.~G.~Makris, R.~El-Ganainy, D.~N.~Christodoulides, and Z.~H.~Musslimani,
``Beam Dynamics in $\mathcal{P}\mathcal{T}$ Symmetric Optical Lattices''
\doi{10.1103/physrevlett.100.103904 } {Phys. Rev. Lett. {\bf 100} (2008), 103904}

\bibitem{longhi}
S.~Longhi,
``Optical Realization of Relativistic Non-Hermitian Quantum Mechanics,''
\doi{10.1103/PhysRevLett.105.013903} {Phys. Rev. Lett. \textbf{105}, (2010) 013903}



\bibitem{Ashida:2017rso}
Y.~Ashida, S.~Furukawa and M.~Ueda,
``Parity-time-symmetric quantum critical phenomena,''
\doi{10.1038/ncomms15791} {Nature Commun. \textbf{8} (2017) no.1, 15791}
 
\bibitem{El-Ganainy:2018ksn}
R.~El-Ganainy, K.~G.~Makris, M.~Khajavikhan, Z.~H.~Musslimani, S.~Rotter and D.~N.~Christodoulides,
``Non-Hermitian physics and PT symmetry,''
\doi{10.1038/nphys4323} {Nature Phys. \textbf{14} (2018) no.1, 11-19}

\bibitem{Midya_2018}
B.~Midya, H. Zhao, and L. Feng,
``Non-Hermitian photonics promises exceptional topology of light''
\doi{10.1038/s41467-018-05175-8} {Nature Communications {\bf 9} (2018), 2674 }

\bibitem{Matsumoto:2019are}
N.~Matsumoto, K.~Kawabata, Y.~Ashida, S.~Furukawa and M.~Ueda,
``Continuous Phase Transition without Gap Closing in Non-Hermitian Quantum Many-Body Systems,''
\doi{10.1103/PhysRevLett.125.260601} {Phys. Rev. Lett. \textbf{125} (2020) no.26, 260601}
\arXiv{1912.09045}{cond-mat.stat-mech}. 

\bibitem{Ashida:2020dkc}
Y.~Ashida, Z.~Gong and M.~Ueda,
``Non-Hermitian physics,''
\doi{10.1080/00018732.2021.1876991} {Adv. Phys. \textbf{69} (2021) no.3, 249-435}
\arXiv{2006.01837}{cond-mat.mes-hall}. 

\bibitem{Rodrigo} R. Bufalo and N. B. Xavier,
``On the presence of electric dipole and toroidal moments in non-Hermitian QED'',
\doi{10.1016/j.physletb.2024.139167} {Phys. Lett. B {\bf 860}, 139167 (2025).}

\bibitem{Cabral:2025mjf}
D.~S.~Cabral, A.~F.~Santos, R.~Bufalo and N.~B.~Xavier,
``Non-Hermitian electron-positron annihilation under thermal effects,''
\doi{10.1140/epjp/s13360-025-06229-9}{Eur. Phys. J. Plus \textbf{140}, 279 (2025)},
\arXiv{2503.15155}{hep-ph}


\bibitem{afsantos1} A. F. Santos and F. C. Khanna, ``Lorentz violation in Bhabha scattering at finite temperature,'' \doi{10.1103/PhysRevD.95.125012}{{Phys. Rev. D} {\bf 95}, 125012 (2017)}.

\bibitem{afsantos2} A. F. Santos and F. C. Khanna, ``Lorentz violation, gravitoelectromagnetism and Bhabha scattering at finite temperature,'' \doi{10.1142/S021773231850061X}{{Modern Physics Letters A} {\bf 33}, 1850061 (2018)}.

\bibitem{afsantos3} A. F. Santos and F. C. Khanna, ``Bhabha scattering in very special relativity at finite temperature,'' \doi{10.1140/epjc/s10052-020-8290-2}{{The European Physical Journal C} {\bf 80}, 703 (2020)}.


\bibitem{khannatfd}  F.~C.~Khanna, A.~P.~C.~Malbouisson, J.~M.~C.~Malbouisson and A.~R.~Santana,
``Thermal quantum field theory - Algebraic aspects and applications,''
World Scientific Publishing Company (2009).

\bibitem{temp00} T. Matsubara, ``A New Approach to Quantum-Statistical Mechanics,'' \doi{10.1143/PTP.14.351}{{Progress of Theoretical Physics} {\bf 14}, 4 (1955)}.

\bibitem{temp1} L. Dolan and R. Jackiw, ``Symmetry behavior at finite temperature,'' \doi{10.1103/PhysRevD.9.3320} {{Phys. Rev. D} {\bf 9}, 3320 (1974)}.

\bibitem{temp2} A. Das and A. Karev, ``Derivative expansion and the chiral anomaly at finite temperature,'' \doi{10.1103/PhysRevD.36.623}{{Phys. Rev. D} {\bf 36}, 623 (1987)}.

\bibitem{temp3} A. Das and J. Frenkel, ``Infrared chiral anomaly at finite temperature,'' \doi{10.1016/j.physletb.2011.01.017}{{Phys. Lett. B} {\bf 696}, 5 (2011)}.

\bibitem{temp000} H. Umezawa, H. Matsumoto and M. Tachiki. ``Thermo field dynamics and condensed states'', Netherlands (1982).

\bibitem{scatter2} D. S. Cabral and A. F. Santos, ``Compton scattering in TFD formalism,'' \doi{10.1140/epjc/s10052-023-11182-x}{{Eur. Phys. J. C} {\bf 83}, 25 (2023)}.


\bibitem{Ohlsson:2019noy}
T.~Ohlsson and S.~Zhou,
``Transition Probabilities in the Two-Level Quantum System with PT-Symmetric Non-Hermitian Hamiltonians,''
\doi{10.1063/5.0002958}{J. Math. Phys. \textbf{61} (2020) no.5, 052104}
\arXiv{1906.01567}{quant-ph}.

\bibitem{Alexandre:2020gah}
J.~Alexandre, J.~Ellis and P.~Millington,
``Discrete spacetime symmetries and particle mixing in non-Hermitian scalar quantum field theories,''
\doi{10.1103/PhysRevD.102.125030}{Phys. Rev. D \textbf{102} (2020) no.12, 125030}
\arXiv{2006.06656}{hep-th}.


\bibitem{regdelta} N. P. Landasman and Ch. G. Van Weert, ``Real- and imaginary-time field theory at finite temperature and density,'' \doi{10.1016/0370-1573(87)90121-9}{{Phys. Rep.} {\bf 145}, 3 (1987)}.

\bibitem{29gev} M. Derrick et. al. ``Experimental study of the reactions $e^+e^-\to e^+e^-$ and $e^+e^-\to\gamma\gamma$ at 29 GeV,'' \doi{10.1103/PhysRevD.34.3286}{{Phys. Rev. D} {\bf 34}, 3286 (1986)}.

\bibitem{43gev} TASSO Collaboration et. al. ``A study of Bhabha scattering at PETRA energies,'' \doi{10.1007/BF01579904}{{Z. Phys. C. - Particles and Fields} {\bf 37}, 171 (1988)}.


\bibitem{Mostafazadeh:2008pw}
A.~Mostafazadeh,
``Pseudo-Hermitian Representation of Quantum Mechanics,''
\doi{10.1142/S0219887810004816}{Int. J. Geom. Meth. Mod. Phys. \textbf{7} (2010), 1191-1306}
\arXiv{0810.5643}{quant-ph}.

\bibitem{Alexandre:2018uol}
J.~Alexandre, J.~Ellis, P.~Millington and D.~Seynaeve,
``Spontaneous symmetry breaking and the Goldstone theorem in non-Hermitian field theories,''
\doi{10.1103/PhysRevD.98.045001}{Phys. Rev. D \textbf{98} (2018), 045001}
\arXiv{1805.06380}{hep-th}.

\bibitem{Mannheim:2018dur}
P.~D.~Mannheim,
``Goldstone bosons and the Englert-Brout-Higgs mechanism in non-Hermitian theories,''
\doi{10.1103/PhysRevD.99.045006}{Phys. Rev. D \textbf{99} (2019) no.4, 045006}
\arXiv{1808.00437}{hep-th}.


\bibitem{Alexandre:2019jdb}
J.~Alexandre, J.~Ellis, P.~Millington and D.~Seynaeve,
``Spontaneously Breaking Non-Abelian Gauge Symmetry in Non-Hermitian Field Theories,''
\doi{10.1103/PhysRevD.101.035008}{Phys. Rev. D \textbf{101} (2020) no.3, 035008}
\arXiv{1910.03985}{hep-th}.



\end{thebibliography}
\end{document}